\documentclass[aps,pre,reprint,onecolumn]{revtex4-2}
\def\rtx@apsprf{
\class@info{APS journal PRF selected}
}

\usepackage{svg}
\usepackage{float}
\usepackage{soul} 
\usepackage{amsmath}
\usepackage{amssymb}
\usepackage[font=small, labelfont=bf, justification=justified]{caption}
\usepackage{subfig}
\usepackage{graphicx}
\usepackage{dcolumn}
\usepackage{bm}
\usepackage{hyperref}
\usepackage{booktabs}

\hypersetup{colorlinks=true, citecolor=blue, linkcolor=blue}
\DeclareMathAlphabet{\altmathcal}{OMS}{cmsy}{m}{n}

\begin{document}

\title{Waviness and self-sustained turbulence in plane Couette-Poiseuille flow}

\author{M. Etchevest}
\affiliation{Universidad de Buenos Aires, Facultad de Ciencias Exactas y Naturales, Departamento de Física, Ciudad Universitaria, 1428 Buenos Aires, Argentina}
\affiliation{CONICET - Universidad de Buenos Aires, Instituto de Física Interdisciplinaria y Aplicada (INFINA), Ciudad Universitaria, 1428 Buenos Aires, Argentina}
\affiliation{Institut Franco-Argentin de Dynamique des Fluides pour l'Environnement, IRL 2027, CNRS, Universidad de Buenos Aires, CONICET. Buenos Aires, Argentina}

\author{P. Dmitruk}
\affiliation{Universidad de Buenos Aires, Facultad de Ciencias Exactas y Naturales, Departamento de Física, Ciudad Universitaria, 1428 Buenos Aires, Argentina}
\affiliation{CONICET - Universidad de Buenos Aires, Instituto de Física Interdisciplinaria y Aplicada (INFINA), Ciudad Universitaria, 1428 Buenos Aires, Argentina}
\affiliation{Institut Franco-Argentin de Dynamique des Fluides pour l'Environnement, IRL 2027, CNRS, Universidad de Buenos Aires, CONICET. Buenos Aires, Argentina}

\author{S. Karmakar}
\affiliation{Université Savoie Mont Blanc, CNRS UMR 5271, LOCIE, 73376 Le Bourget du Lac, France}

\author{B. Semin}
\affiliation{Physique et Mécanique des Milieux Hétérogènes (PMMH UMR 7636)\\ CNRS, ESPCI Paris, Université PSL, Sorbonne Université, Université Paris Cité, Paris, France}

\author{R. Godoy-Diana}
\affiliation{Physique et Mécanique des Milieux Hétérogènes (PMMH UMR 7636)\\ CNRS, ESPCI Paris, Université PSL, Sorbonne Université, Université Paris Cité, Paris, France}

\author{J.E. Wesfreid}
\affiliation{Physique et Mécanique des Milieux Hétérogènes (PMMH UMR 7636)\\ CNRS, ESPCI Paris, Université PSL, Sorbonne Université, Université Paris Cité, Paris, France}

\begin{abstract}

Direct numerical simulations of a Couette–Poiseuille flow were performed near the transition to turbulence to investigate the nonlinear relationship between streak waviness and rolls. This relationship is a key step in Waleffe's model for a self-sustaining process (SSP). Simulations were conducted for Reynolds numbers ranging from 500 to 940, and a range of initial perturbation amplitudes was used.
In these simulations, the streaks, rolls, and streak waviness initially grow. The optimal time for this growth closely matches the linear transient growth period for small perturbations, but is much shorter when the initial perturbations are large and highly nonlinear. For higher Reynolds numbers and large initial perturbations, the velocity field reaches a turbulent steady state, while in the remaining cases the flow relaminarizes. The main result is that the waviness of the streaks is a quadratic function of the rolls, provided that the roll amplitude is sufficiently large.

\end{abstract}

\maketitle

\section{Introduction}\label{Introduction}

\noindent Shear flows are ubiquitous in natural and industrial environments, from atmospheric and oceanic currents to engineering applications such as pipeline transport and vehicle aerodynamics. Understanding the transition to turbulence in these flows is essential for improving energy efficiency, developing control strategies, and predicting flow behavior under a wide range of conditions. From a fundamental perspective, the study of turbulence and the mechanisms governing the transition remains an area with many unresolved questions.

Streamwise streaks and rolls are widely recognized as the fundamental building blocks of the transition to turbulence in wall-bounded shear flows \cite{elder1960experimental,klebanoff1962three,kline1967structure}. 
They play a central role in both the onset of turbulence and its self-sustained maintenance, even at high Reynolds numbers.
Above a critical Reynolds number, the turbulent regime can be achieved and maintained through a mechanism known as the self-sustaining process (SSP) \cite{jimenez1991minimal,hamilton1995regeneration,waleffe1997}. This mechanism involves a closed loop of energy transfer between streaks and rolls, driven by a combination of linear and nonlinear interactions.

The formation of elongated streaks arises from the lift-up effect, where streamwise vortices advect fluid with different velocities away from the walls, producing a striped streak pattern \cite{landahl1975wave}. These streaks can develop wavy instabilities caused by the shear between adjacent fast and slow streaks \cite{hamilton1995regeneration,waleffe1997,farrell2012dynamics}. Eventually, the instability can lead to streak breakdown, transferring energy back to the rolls through nonlinear interactions. This process has been investigated both experimentally and numerically in boundary layers and channel flows\cite{mans2005breakdown,mans2007sinuous,brandt2004transition,cossu2017self,duriez2009self} and in centrifugally unstable configurations \cite{swearingen1987growth,martinand2014mechanisms,dessup2018self}.

A canonical example of wall-bounded shear flows is plane Couette flow (PCF), where a channel flow is driven by the motion of one or both walls. PCF is linearly stable, with an infinite critical Reynolds number $Re_c$ \cite{davey1973stability}. Because of its simplicity and well-defined boundary conditions, PCF has long served as a model flow to investigate fundamental transition mechanisms such as the SSP.

Numerous experimental observations in shear flows such as PCF, plane Poiseuille flow (PPF), and pipe flow reveal that turbulence can arise at Reynolds numbers far below the linear stability threshold. This subcritical transition to turbulence \cite{MANNEVILLE2016} originates from finite-amplitude perturbations and involves transient growth of disturbances \cite{schmid2012stability} and the emergence of exact coherent structures (ECS), including edge states that separate laminar and turbulent regimes in phase space \cite{duguet_schlatter_henningson2009}. In this regime, pattern formation leads to the coexistence of turbulent and laminar patches, a topic of active current research \cite{benavides2025, ciola2025large,gome2023patterns_part1}.

\begin{figure}[H]
    \centering
    \includegraphics[width=0.8\linewidth]{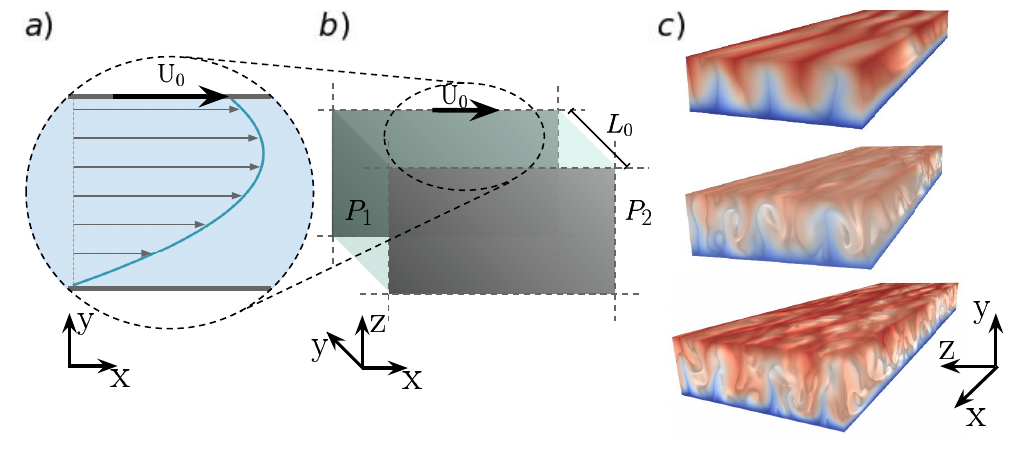}
    \caption{(a,b) Schematic representation of a typical Couette-Poiseuille laminar flow considered. (c) Representative 3D snapshots of  the streamwise velocity field $U_x$ (blue is low speed, red is high speed) for three cases, a decaying laminar flow at $Re=625$, a transitioning flow at $Re=806$, and a fully turbulent case at $Re=940$, from top to bottom, respectively. Rendering using the software VAPOR \cite{li2019vapor,sgpearse_2023_vapor}. }
    \label{Fig1_setup}
\end{figure}

Coupling plane Poiseuille and plane Couette flows yields plane Couette–Poiseuille flow (CPF), shown schematically in Fig.~\ref{Fig1_setup}. CPF is linearly stable under certain conditions, such as when the wall velocity ($U_0$) exceeds $70\%$ of the centerline velocity \cite{Potter_1966}. Its transitional dynamics have been observed experimentally \cite{klotz2017cpf_exp,Liu_Semin_Klotz_Godoy-Diana_Wesfreid_Mullin_2021,Liu2024} and numerically \cite{KIM2018,Kashyap2020,Cheng_Pullin_Samtaney_Luo_2023}, revealing a rich interplay between laminar, transitional, and turbulent regimes. This makes CPF a particularly suitable configuration for examining the nonlinear interaction between streaks and rolls near the transition to turbulence using direct numerical simulations (DNS).

\begin{figure}[h]
    \centering
    \includegraphics[scale=0.8]{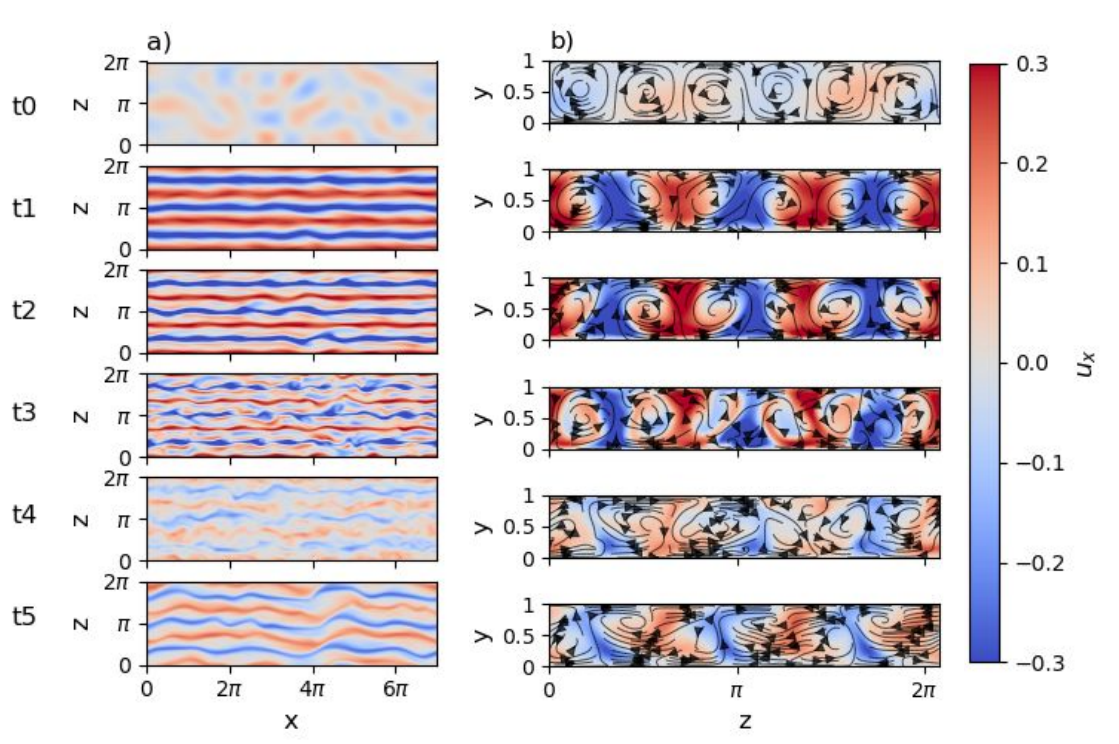}
    \caption{(a,b) Snapshots of a typical flow field at $Re=806$ (Simulation T16 in Table \ref{table1} from Section \ref{Numerical set up}) at 5 different times $t0=0$, $t1=4$, $t2=7$, $t3=11$, $t4=35$ and $t5=100$, with time nondimensionalized by $L_0/U_0$. In (a) $xz$-plane at fixed $y = L_y/2$, and in (b) the $yz$-plane at fixed $x = L_x/2$ are presented. The heatmap in the two panels represent the fluctuations of the streamwise velocity component $u_x$, and the streamplot in panel (b) corresponds with the cross-stream flow fields.}
    \label{Fig2_streaks}
\end{figure}

\begin{figure}[h]
    \centering
    \includegraphics[scale=0.65]{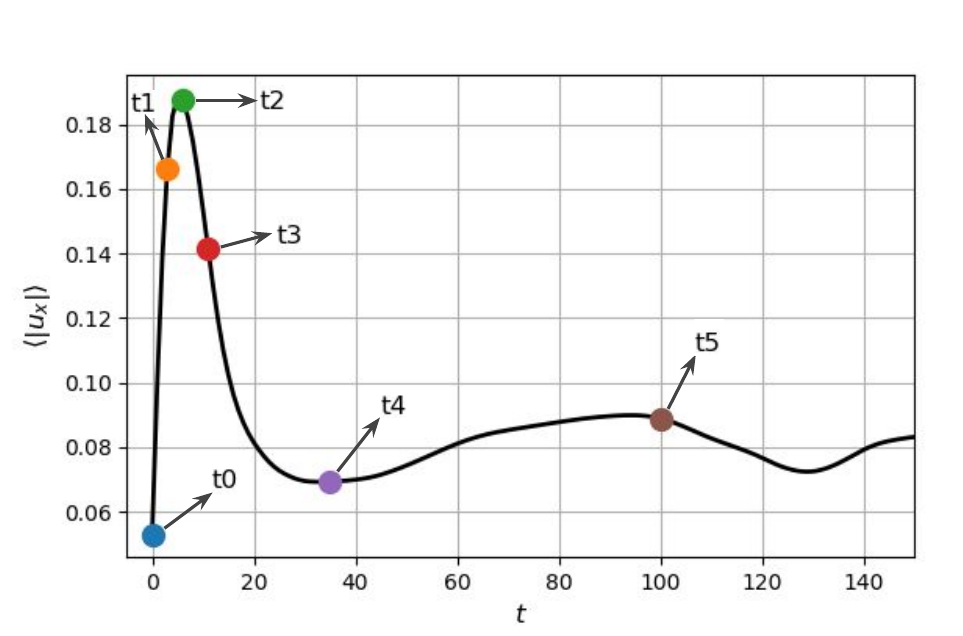}
    \caption{Streak amplitude $\langle |u_x| \rangle$ evolution as a function of time of the simulation T16 (see Table \ref{table1}). The times corresponding to the snapshots in Fig.~\ref{Fig2_streaks} (a) and (b) are indicated along the curve.}
    \label{Fig3_uxtime}
\end{figure}

Figure~\ref{Fig2_streaks} illustrates the temporal evolution of the streaks and rolls for a representative case in which the streaks grow from a noisy perturbation with imposed streamwise rolls, eventually developing a wavy instability.
Panel (a) shows snapshots of the streamwise velocity at times $t_0=0$, $t_1=4$, $t_2=7$, $t_3=11$, $t_4=35$ and $t_5=100$, where time is expressed in units of $L_0/U_0$ in the $x$–$y$ plane, corresponding to simulation T16 at $Re=806$ (see Table \ref{table1} in Section \ref{Numerical set up}). Panel (b) displays the same evolution in the $y$–$z$ plane, with cross-stream velocity superimposed. The streak intensity $\langle |u_x|\rangle$ as defined in Eq.~\ref{u_x} of Section~\ref{Numerical set up} is shown as a function of time in Fig.~\ref{Fig3_uxtime}.

Initially, streaks grow from a noisy perturbation with imposed streamwise rolls (panels $t_0$–$t_1$). As high-velocity fluid is advected downward and low-velocity fluid upward, a characteristic up–down advection pattern develops, analogous to the plume structures in Rayleigh–Bénard convection. This process corresponds to the increasing streak energy shown in Fig. \ref{Fig3_uxtime}, that reaches a maximum at $t_2$. The streak waviness becomes evident at $t_2$, followed by streak breakdown at $t_3$–$t_4$, when the cross-stream structures decay and turbulence develops. Finally, the cycle closes with regenerated streaks and rolls at $t_5$, corresponding to a secondary peak in streak energy.

Nonlinear minimal models of this SSP have been successfully obtained from truncated representations of the Navier–Stokes equations \cite{waleffe1997,moehlis2004low,manneville2018,cavalieri2022}. Other models extend this framework to include spatial inhomogeneity \cite{Manneville_2012,kashyap2025laminar} and large-scale flows \cite{benavides2025}. One of the simplest is Waleffe’s fourth-order model \cite{waleffe1997}, which describes the coupled dynamics of streaks, rolls, wall-normal vorticity, and mean flow distortion. Previous experiments have measured the lift-up effect in laminar CPF \cite{Liu2024}, showing a linear relationship between streak and roll amplitudes. As streak waviness increases, the system departs from this linear regime, a behavior captured by one of Waleffe’s equations (Eq. (20) in \cite{waleffe1997}, corresponding to Eq.~\ref{U_waleffe} here). This motivates further investigation of the nonlinear feedback mechanisms linking streak waviness and roll amplification, which is the focus of the present paper.

This work analyzes the energy feedback between streaks and rolls near the transition to turbulence in CPF, focusing on the nonlinear mechanisms triggered by streak waviness. We also assess the consistency of this behavior with Waleffe’s truncated model.

The manuscript is organized as follows: Section \ref{Numerical set up} presents the numerical setup and governing equations; Section \ref{Results} reports the main results and compares them with previous theoretical work; Section \ref{Conclusions} summarizes the findings and outlines perspectives for future research.

\section{Numerical set up}\label{Numerical set up}

\noindent We consider an incompressible viscous fluid confined between two parallel plates. One of the plates moves at a constant velocity \( U_0 \) in the streamwise direction, while a constant mean pressure gradient \( (P_2 - P_1)/L_x \) is imposed along the same direction. This configuration corresponds to a plane Couette–Poiseuille flow (CPF). The computational domain is rectangular, with dimensions \( L_x \times L_y \times L_z = (7\pi \times 1 \times 2\pi) L_0 \), where $L_0$ is the channel width and serves as the reference length scale.
Periodic boundary conditions are applied in the streamwise (\( \hat{x} \)) and spanwise (\( \hat{z} \)) directions, and no-slip impermeable conditions are imposed on the walls in the wall-normal (\( \hat{y} \)) direction. A diagram of the configuration and a representative laminar velocity profile are shown in Fig.~\ref{Fig1_setup}(a).

The flow dynamics are governed by the incompressible Navier–Stokes equations in nondimensional form,

\begin{align}
    \frac{\partial \boldsymbol{U}}{\partial t} + (\boldsymbol{U} \cdot \boldsymbol{\nabla}) \boldsymbol{U} &= -\boldsymbol{\nabla} P + \frac{1}{2Re} \nabla^2 \boldsymbol{U}, \label{HD} \\
    \boldsymbol{\nabla} \cdot \boldsymbol{U} &= 0, \label{incompressibility}
\end{align}
where \( \boldsymbol{U} \) is the velocity field in units of the wall velocity $U_0$, \( P \) the pressure field nondimensionalized by $U_0^2$, and \( Re \) the Reynolds number. The fluid density is set to unity, and a constant mean pressure gradient is imposed, given by
\begin{equation}
    f_0 = \frac{(P_2 - P_1)/L_x}{L_0/U_0^2},
\end{equation}
The Reynolds number is defined as
\begin{equation}
    Re = \frac{U_0 L_0}{2 \nu},
\end{equation}
\noindent with the characteristic length scale taken as the channel half-width \( L_0/2 \). This choice facilitates direct comparison with previous studies.

Theoretical analyses of similar flows have employed reduced-order models to describe the dynamics of coherent structures \cite{waleffe1997,moehlis2004low,manneville2018,cavalieri2022}. In particular, the four-equation model proposed by Waleffe \cite{waleffe1997} captures the essential feedback mechanisms between streaks, rolls, and waviness in plane Couette flow:
\begin{align}
    &\left(\frac{d}{dt} + \frac{\kappa_m^2}{Re}\right) M = \sigma_m W^2 - \sigma_u U V + \frac{\kappa_m^2}{Re}, \label{M_waleffe}\\ 
    &\left(\frac{d}{dt} + \frac{\kappa_u^2}{Re}\right) U = -\sigma_w W^2 + \sigma_u M V, \label{U_waleffe}\\
    &\left(\frac{d}{dt} + \frac{\kappa_v^2}{Re}\right) V = \sigma_v W^2, \label{V_waleffe}\\
    &\left(\frac{d}{dt} + \frac{\kappa_w^2}{Re}\right) W = \sigma_w U W - \sigma_m M W - \sigma_v V W. \label{W_waleffe}
\end{align}
Here, \( M \), \( U \), \( V \), and \( W \) denote the amplitudes of the mean profile distortion, streaks, rolls, and streak waviness, respectively. The parameters \( \kappa_i \) are the characteristic wavenumbers and \( \sigma_i \) the nonlinear coupling coefficients. Although this model was originally formulated for plane Couette flow, we assume that its essential mechanisms remain qualitatively valid in CPF and provide a conceptual framework to interpret our numerical results. This assumption is supported by recent experimental measurements of the lift-up effect in CPF \cite{Liu2024}, which show good agreement with model predictions.

The DNS were carried out with SPECTER \cite{FONTANA2020}, a pseudospectral solver employing a Fourier basis in the periodic directions and Fourier continuation methods in the wall-normal direction \cite{Bruno20102009}.

For the analysis of streak dynamics, the velocity field is decomposed as
\begin{equation}
    \boldsymbol{U} = \langle \boldsymbol{U} \rangle + \boldsymbol{u},
\end{equation}
where \( \langle \boldsymbol{U} \rangle \) is the mean flow and \( \boldsymbol{u} \) the fluctuations. The mean flow has only a streamwise component, \( \langle \boldsymbol{U} \rangle = \langle U_x \rangle_{x,z}(y) \), and the streaks are characterized by \( u_x \). Here, $U_x$ denotes the streamwise component of the velocity field, and the notation $\langle \cdot \rangle_x$, $\langle \cdot \rangle_y$, and $\langle \cdot \rangle_z$ denotes averages over the streamwise, wall-normal, and spanwise directions, respectively. Combinations of subscripts, such as $\langle \cdot \rangle_{x,z}$, indicate averaging over the corresponding set of directions.

\begin{figure}[H]
    \centering
    \includegraphics[scale=0.7]{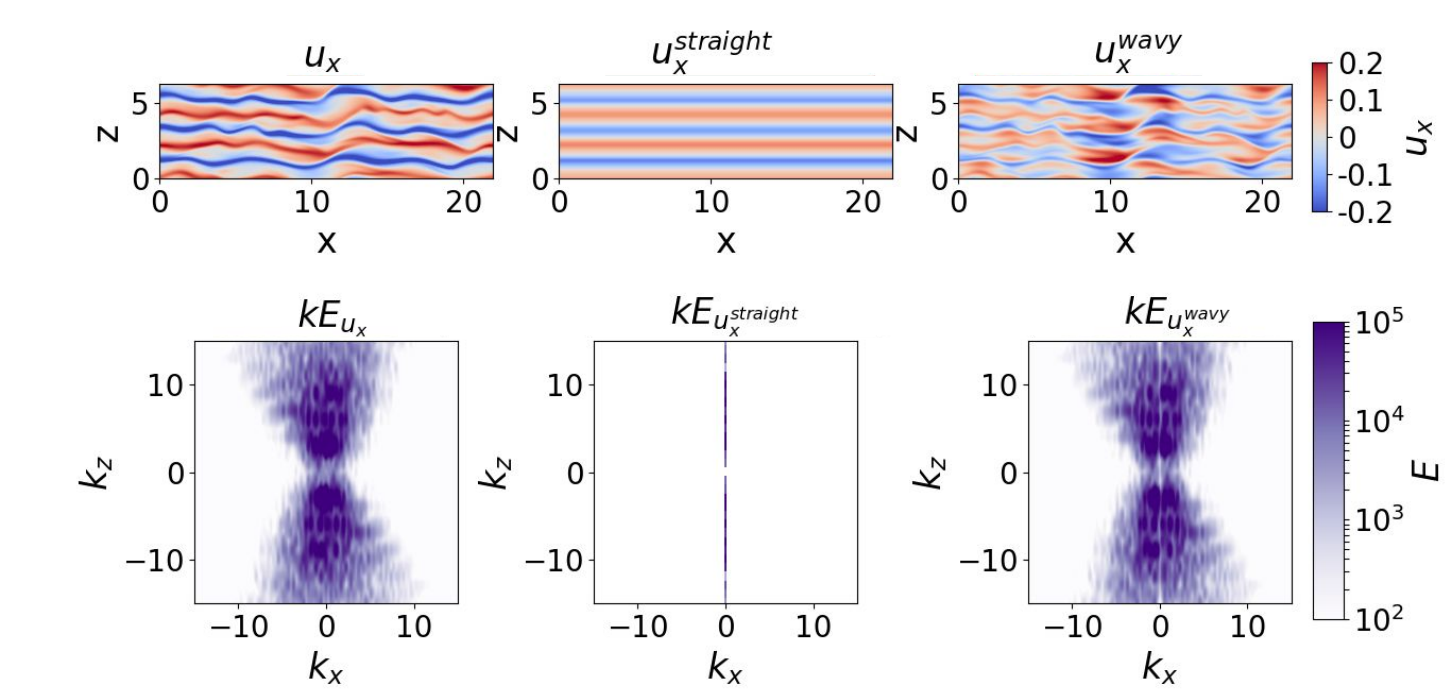}
    \caption{In the upper panels, the streamwise velocity field $u_x$ (left) and its decomposition into the straight part (middle) and wavy part (right) of a simulation with $Re=806$ (run T16), shown at fixed $y = L_y/2$, are displayed. Additionally, the power spectra of each field are presented in the lower panels.}
    \label{Fig4_filtering}
\end{figure}

To characterize the nonlinear regeneration of streamwise rolls induced by streak waviness, we define the following modal amplitudes:
\begin{align}
     \langle |u_x| \rangle &= \frac{1}{\mathcal{V}}\int_\mathcal{V} |u_x| \, d\mathcal{V}, \label{u_x}\\
     \langle |u_y| \rangle &= \frac{1}{\mathcal{V}}\int_\mathcal{V} |u_y| \, d\mathcal{V}, \label{u_y}\\
     \langle |\omega_y^{wavy}| \rangle &= \frac{1}{\mathcal{V}}\int_\mathcal{V} \left| \frac{\partial u_x^{wavy}}{\partial z} - \frac{\partial u_z}{\partial x} \right| d\mathcal{V}, \label{w_y}\\
     m &= \frac{1}{\mathcal{V}}\frac{\int_\mathcal{V} |U_x| \, d\mathcal{V}}{\langle |U_x^{lam}(y)| \rangle_y}, \label{M}
\end{align}
where \( U_x^{lam}(y) = f_0 Re\, y^2 + \left( 1 - f_0 Re \right)\, y \) is the theoretical laminar profile and $\mathcal{V}$ denotes the volume of the computational domain. The wavy part of the streamwise velocity, \( u_x^{wavy} \), is obtained by subtracting the straight component (the \( k_x = 0 \) Fourier mode, i.e., \( \langle u_x \rangle_x \)) from the total field. Figure~\ref{Fig4_filtering} illustrates this decomposition, with the corresponding power spectra shown in the lower panels. The quantities $\langle |u_x| \rangle$, $\langle |u_y| \rangle$, $\langle |\omega_y^{wavy}| \rangle$ and $m$ are used as proxies for the variables $U$, $V$, $W$ and $M$ of Waleffe’s reduced model, respectively. Note that, although $m$ is defined here for completeness, it will not be discussed further in the present work.

A total of 39 simulations at different Reynolds numbers were analyzed using various initial conditions (see Table~\ref{table1}). Two types of initial conditions were employed. In the first case, the laminar base flow was perturbed with three-dimensional incompressible random fluctuations. These perturbations consist of white noise restricted to a spectral shell in the streamwise and spanwise directions, with wavenumbers $k_x$ and $k_z$ between 2 and 5. In the wall-normal direction, the perturbations satisfy the boundary conditions using the Chandrasekhar–Reid harmonics, following the formulation described in \cite{CLEVER_BUSSE_1997}, also restricted to the same spectral shell. In the second case, the same laminar base flow perturbed with random perturbations was used as a background field, but additional streamwise rolls were superimposed. These rolls were introduced analytically through the wall-normal and spanwise velocity components, defined as
\[
u_y = \frac{1}{5}\sin(\pi y)\cos(3 z)\Gamma(y), \qquad 
u_z = -\frac{1}{5}\cos(\pi y)\sin(3 z)\Gamma(y),
\]
which generate three pairs of counter-rotating streamwise vortices and $\Gamma(y)$ is a \textit{Tukey} window to enforce the boundary conditions. This construction allows us to control the initial roll intensity while preserving the structure of the laminar base flow. Reynolds numbers in the range \( 500 \leq Re \leq 940 \) were explored, with multiple realizations for each \( Re \) to vary both the perturbation structure and its amplitude. The corresponding parameters are listed in Table~\ref{table1}. Both types of initial conditions were tested, but no notable differences were found. Therefore the simulations in Table~\ref{table1} are reported without specifying the shape of the initial condition.

\begin{table}[h]
\begin{center}
\begin{tabular*}{.9\textwidth}{ c @{\extracolsep{\fill}} c c c c c}
\toprule
\toprule
ID & $Re$ & $\langle|u_x|\rangle(t=0)$ & $\langle|u_y|\rangle(t=0)$ & $\langle|\omega_y^{wavy}|\rangle(t=0)$ & $LP$ \\ \hline
\midrule

L1 & $625$ & $0.016$ & $0.007$ & $0.055$ & $ 1.625 $ \\ 
L2 & $800$ & $0.012$ & $0.004$ & $0.031$ & $ 1.8 $ \\ 
L3 & $806$ & $0.012$ & $0.004$ & $0.031$ & $ 1.806 $  \\ 
L4 & $806$ & $0.005$ & $0.002$ & $0.014$ & $ 1.806 $  \\ 
L5 & $806$ & $0.012$ & $0.005$ & $0.039$ & $ 1.806 $  \\ 
L6 & $806$ & $0.005$ & $0.002$ & $0.018$ & $ 1.806 $ \\ 

NL1 & $500$ & $0.034$ & $0.021$ & $0.176$ & $ 1.5 $  \\ 
NL2 & $625$ & $0.035$ & $0.022$ & $0.178$ & $ 1.625 $  \\ 
NL3 & $625$ & $0.043$ & $0.026$ & $0.219$ & $ 1.625 $ \\ 
NL4 & $625$ & $0.036$ & $0.016$ & $0.123$ & $ 1.625 $  \\ 
NL5 & $714$ & $0.033$ & $0.018$ & $0.214$ & $ 1.714 $  \\ 
NL6 & $500$ & $0.073$ & $0.058$ & $0.270$ & $ 1.5 $  \\ 
NL7 & $625$ & $0.056$ & $0.023$ & $0.216$ & $ 1.625 $  \\ 

T1 & $625$ & $0.084$ & $0.062$ & $0.338$ & $ 1.625 $  \\ 
T2 & $800$ & $0.033$ & $0.018$ & $0.113$ & $ 1.8 $  \\ 
T3 & $800$ & $0.053$ & $0.023$ & $0.270$ & $ 1.8 $  \\ 
T4 & $800$ & $0.053$ & $0.054$ & $0.049$ & $ 1.8 $  \\ 
T5 & $806$ & $0.123$ & $0.026$ & $0.163$ & $ 1.806 $  \\ 
T6 & $806$ & $0.135$ & $0.071$ & $0.570$ & $ 1.806 $  \\ 
T7 & $806$ & $0.053$ & $0.054$ & $0.049$ & $ 1.806 $  \\ 
T8 & $940$ & $0.095$ & $0.066$ & $0.402$ & $ 1.94 $  \\ 
T9 & $714$ & $0.035$ & $0.022$ & $0.179$ & $ 1.714 $  \\ 
T10 & $800$ & $0.035$ & $0.022$ & $0.180$ & $ 1.8 $  \\ 
T11 & $800$ & $0.033$ & $0.023$ & $0.226$ & $ 1.8 $  \\ 
T12 & $800$ & $0.053$ & $0.054$ & $0.049$ & $ 1.8 $  \\ 
T13 & $806$ & $0.050$ & $0.031$ & $0.257$ & $ 1.806 $  \\ 
T14 & $806$ & $0.016$ & $0.010$ & $0.080$ & $ 1.806 $ \\ 
T15 & $806$ & $0.053$ & $0.054$ & $0.049$ & $ 1.806 $  \\ 
T16 & $806$ & $0.053$ & $0.054$ & $0.049$ & $ 1.806 $ \\ 
T17 & $806$ & $0.150$ & $0.109$ & $0.950$ & $ 1.806 $ \\ 
T18 & $806$ & $0.016$ & $0.007$ & $0.056$ & $ 1.806 $ \\ 
T19 & $834$ & $0.035$ & $0.022$ & $0.180$ & $ 1.834 $  \\ 
T20 & $834$ & $0.034$ & $0.018$ & $0.221$ & $ 1.834 $ \\ 
T21 & $892$ & $0.035$ & $0.022$ & $0.181$ & $ 1.892 $  \\ 
T22 & $892$ & $0.035$ & $0.010$ & $0.064$ & $ 1.892 $  \\ 
T23 & $892$ & $0.034$ & $0.018$ & $0.224$ & $ 1.892 $ \\ 
T24 & $940$ & $0.050$ & $0.031$ & $0.259$ & $ 1.94 $  \\ 
T25 & $940$ & $0.017$ & $0.008$ & $0.056$ & $ 1.94 $ \\ 
T26 & $940$ & $0.034$ & $0.019$ & $0.226$ & $ 1.94 $ \\ 

\bottomrule
\bottomrule
\end{tabular*}
\end{center}
\caption{Simulations with $256\times103\times128$ grid points, fixed pressure gradient at $f_0=-0.001$, and the wall velocity $U_0$ in $y=L_y$. ID is the label for each simulation, with T, NL, and L denoting the three different behavior identified in this work. T is for turbulent, L for laminar decay, and NL laminar decay with nonlinear fluctuations. $Re$ is the Reynolds number, $\langle|u_x|\rangle(t=0)$, $\langle|u_y|\rangle(t=0)$, and $\langle|\omega_y^{wavy}|\rangle(t=0)$ the initial streaks, rolls, and waviness intensities respectively. Finally linear part refers to the expression $LP=1-f_0 Re$, which defines the coefficient for the linear part of the theoretical laminar profile $U_x^{lam}(y)=f_0 Re\,y^2+ (1-f_0 Re)\, y$. For reference the quadratic part of the expression above is defined by $Q=1-LP$.}
\label{table1}
\end{table}

\section{Results}\label{Results}

\begin{figure}[H]
    \centering
    \includegraphics[scale=0.4]{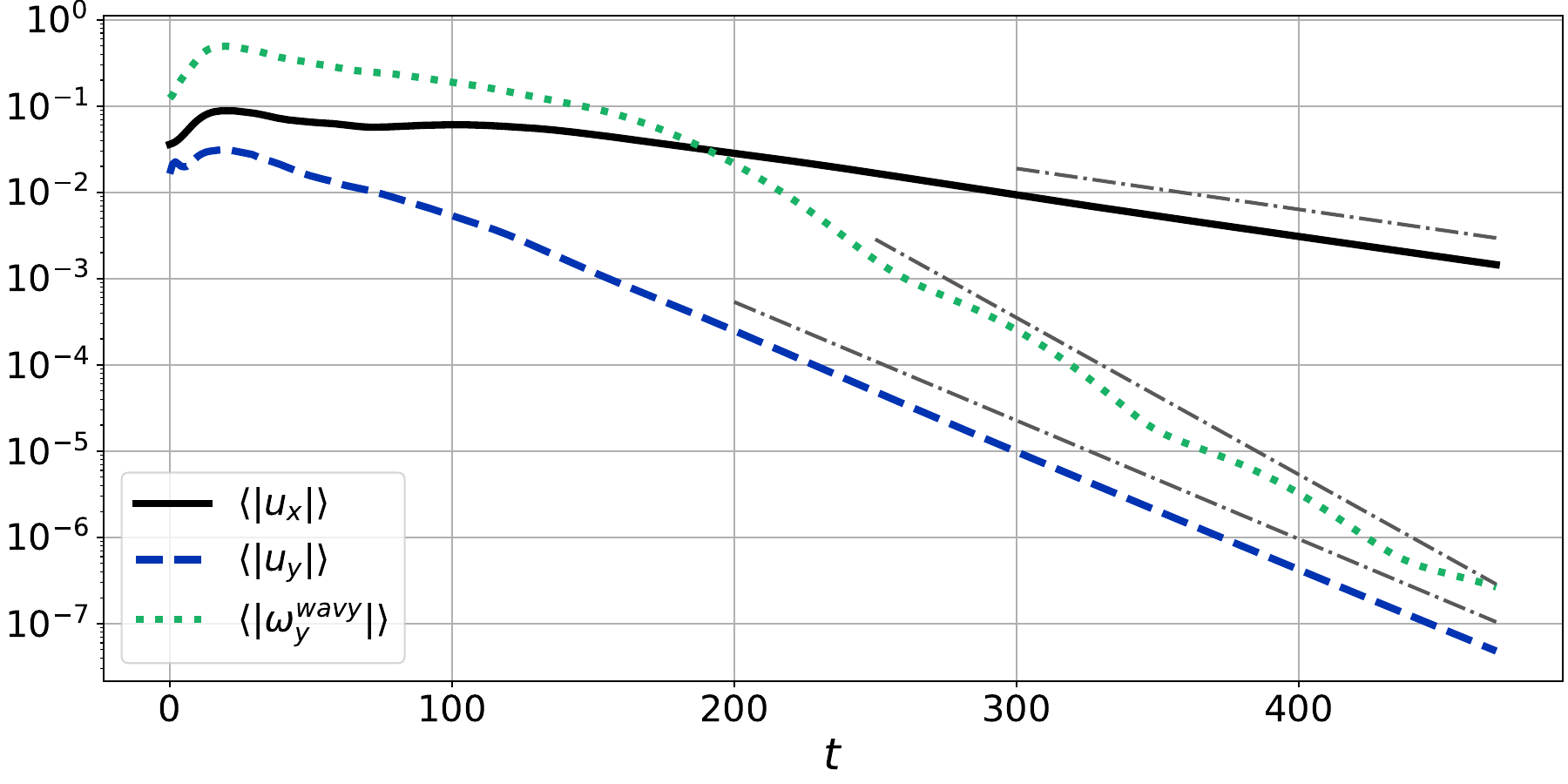}
    \caption{Temporal evolution of $\langle|u_x|\rangle$, $\langle|u_y|\rangle$, and $\langle|\omega_y^{wavy}|\rangle$ for run NL4 ($Re = 625$). The vertical axis is shown in logarithmic scale to highlight the exponential laminar decay, indicated by the three grey dash-dotted lines.}
    \label{Fig5_exponential_decay}
\end{figure}

We begin by examining the laminar behaviour of the flow, which manifests through the decay of fluctuations. In the runs that eventually reach a laminar state (L1–L6 and NL1–NL7), a transient behaviour is observed before an exponential decay sets in. This decaying stage is consistent with laminar relaxation and aligns with Eqs.~(\ref{U_waleffe}), (\ref{V_waleffe}) and (\ref{W_waleffe}), where the exponential decay rate is $\kappa^2/Re$ when the nonlinear terms on the right-hand side are neglected. This rate depends on the characteristic parameter $\kappa$ and the Reynolds number $Re$. Figure~\ref{Fig5_exponential_decay} shows this behaviour for run NL4, which asymptotically reaches a laminar state. The $\kappa$ values are obtained by fitting a straight line to the logarithm of the decaying parts, corresponding to the linearised exponential decay. It is observed that this behaviour appears in all runs, with $\kappa_v$ values remaining nearly constant: $\kappa_v$ lies between $4.33$ and $4.48$. We therefore use the mean value $\kappa_v = 4.4$ for the analysis that follows. It is worth noting that the order of magnitude of $\kappa_v$ matches the observed wavenumber of the rolls and is also consistent with the optimal wavenumber predicted by linear transient growth theory (see Appendix B). In addition, the decay of the rolls ($\langle |u_y|\rangle$) is about $2.0$ to $2.7$ times faster than that of the streaks ($\langle |u_x|\rangle$), consistent with the ratio between $\tau_x$ and $\tau_z$ reported in \cite{Liu_Semin_Klotz_Godoy-Diana_Wesfreid_Mullin_2021}.

Once $\kappa_v$ is estimated, the evolution equation for $\langle |u_y| \rangle$ becomes fully determined except for the coupling coefficient $\sigma_v$, which must be obtained from the data. Comparing the diffusion term $\kappa_v^2/Re \langle |u_y|\rangle$ with the time derivative of $\langle |u_y|\rangle$ shows that the latter remains small in both fluctuating (discussed later) and turbulent states, although it cannot be neglected during the transient or laminar decay stages.

\begin{figure}[t]
    \centering
    \includegraphics[scale=0.5]{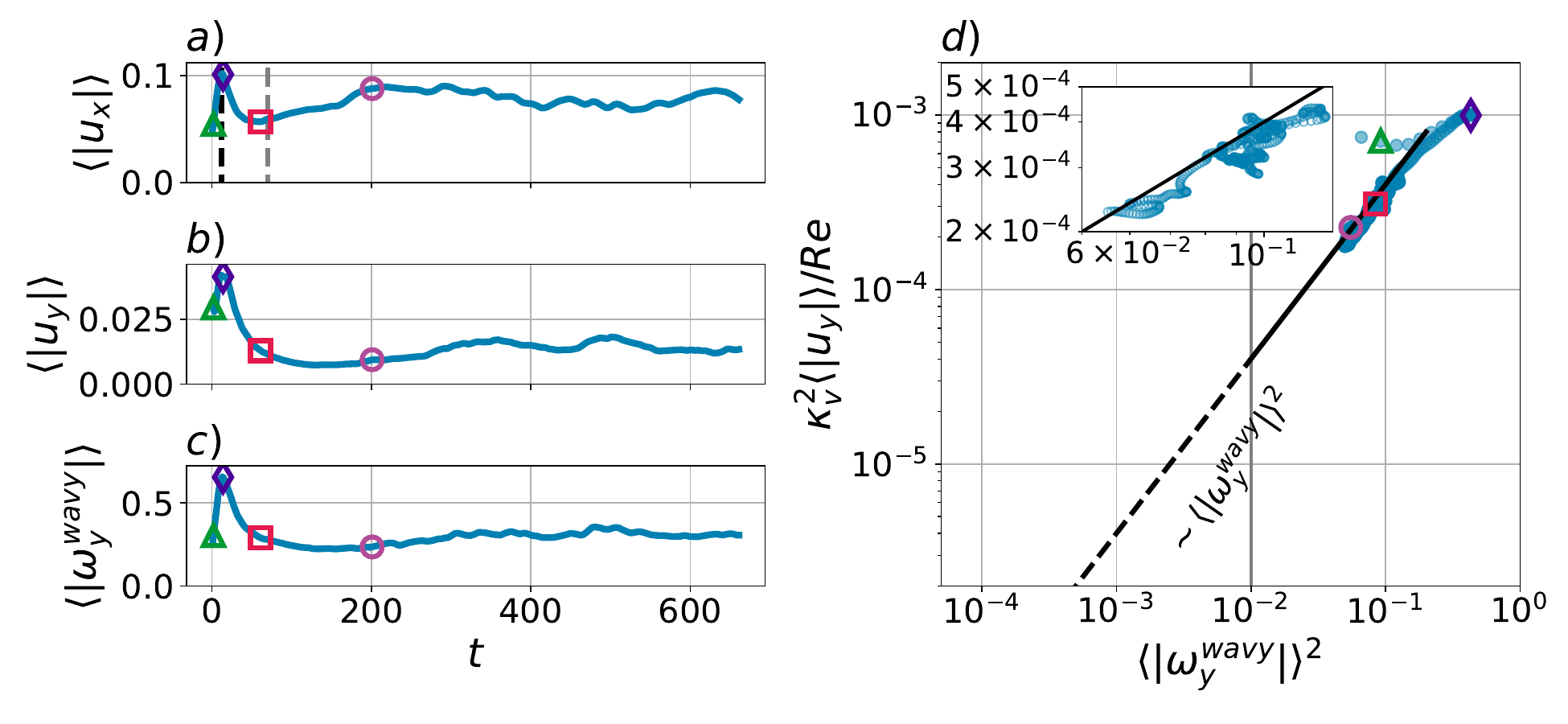}
    \caption{Time evolution of $\langle |u_x|\rangle$, $\langle |u_y|\rangle$, and $\langle |\omega_y^{wavy}|\rangle$ for simulation T13 ($Re = 806$) in panels (a), (b), and (c), respectively. The black and grey dashed lines in panel (a) mark the observed maximum and theoretical optimal linear transient growth times. Panel (d) shows $\kappa_v^2/Re \langle |u_y|\rangle$ vs. $\langle |\omega_y^{wavy}|\rangle^2$ over the entire time range. The black solid line corresponds to the region where the quadratic relationship holds, while the dashed line corresponds to where it does not. The triangle ($t=0$), rhombus ($t=13$), square ($t=60$), and circle ($t=200$) highlight four representative times. The inset isolates the turbulent stage after $t=200$.}
    \label{Fig6_uy_wy2_turb}
\end{figure}

Figure~\ref{Fig6_uy_wy2_turb} illustrates this behaviour for the turbulent simulation T13. Panels (a)–(c) show the time evolution of $\langle |u_x|\rangle$, $\langle |u_y|\rangle$, and $\langle |\omega_y^{wavy}|\rangle$, respectively, while panel (d) presents the diffusion term $\frac{\kappa_v^2}{Re}\langle |u_y|\rangle$ as a function of the squared wall-normal vorticity $\langle |\omega_y^{wavy}|\rangle^2$. 
In panel (a), the observed time of the maximum and the theoretical optimal time of linear transient growth (Eq.~(\ref{eqn:topt}) Appendix \ref{Appendix 2}) are indicated by black and grey dashed lines, respectively; their marked difference reflects a strongly nonlinear regime. 
The pronounced peaks observed are reminiscent of bursting-like events reported in studies of nonlinear optimal perturbations (see, e.g., \cite{Cherubini_DePalma2013}). However, the present work does not aim at characterizing such transient bursting dynamics nor at triggering optimal perturbations. The focus here is instead on the nonlinear coupling between rolls, streaks, and waviness in the turbulent regime.
The flow reaches a turbulent state after the pink circle at $t=200$, this turbulent portion is isolated in panel (d).
A clear quadratic relationship emerges between the dynamics of wall-normal vorticity and the diffusion term, consistent with the nonlinear coupling term in Eq.~(\ref{V_waleffe}), where $\sigma_V$ acts as the proportionality coefficient.
This indicates that the nonlinear interaction between vorticity and rolls is largely captured by this balance.
The same behaviour is observed across all turbulent simulations, with $\sigma_V \approx 4 \times 10^{-3}$, suggesting that a single coupling constant governs the model dynamics and that these nonlinearities are intrinsic to the observed behaviour.

\begin{figure}[h]
    \centering
    \includegraphics[scale=0.5]{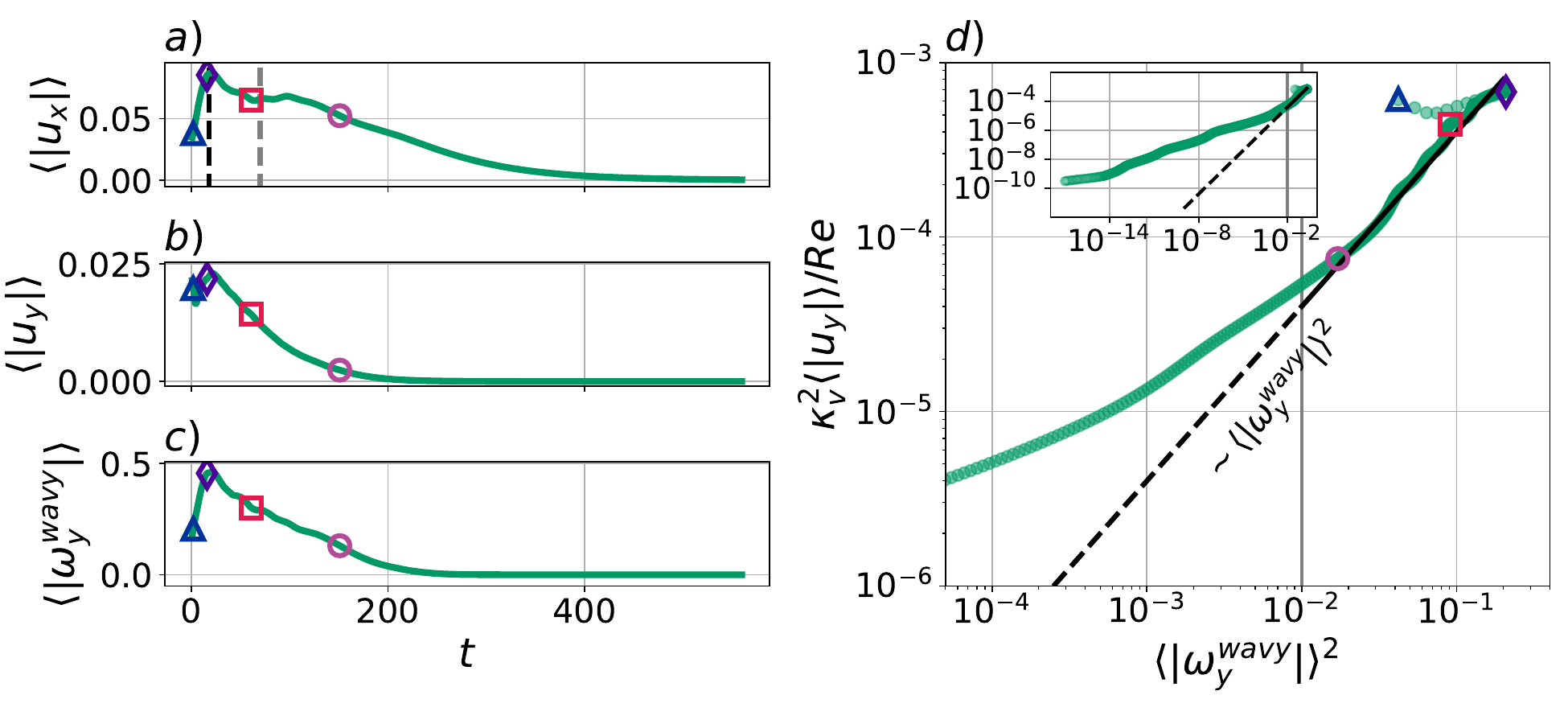}
    \caption{Same as Fig.~\ref{Fig6_uy_wy2_turb}, but for simulation NL2 ($Re = 625$). Only the initial part of the evolution is shown in panel (d), where the $\kappa_v^2/Re \langle |u_y|\rangle \sim \langle |\omega_y^{wavy}|\rangle^2$ relationship holds. The inset in (d) displays the full time series. The marked times are $t=0$, $t=15$, $t=60$, and $t=150$ (triangle, rhombus, square, and circle, respectively).}
    \label{Fig7_uy_wy2_NL}
\end{figure}

In Fig.~\ref{Fig7_uy_wy2_NL}, the same relationship is examined for a decaying simulation (NL2), revealing three distinct regimes. During the initial stage, up to the rhombus point, streak intensity increases (panel a) and both roll and vorticity intensities rise (panels b and c). This clearly signals active nonlinear interactions. Again, the difference between the observed and optimal linear times of maximal gain (black and grey dashed lines in panel a) confirms the nonlinear nature of this phase. 
Different levels of initial perturbation energy were tested in order to explore both situations, with some amplitudes chosen \textit{ad hoc} to trigger nonlinear interactions and others selected to keep the transient predominantly linear. This strategy makes it possible to identify more clearly the onset of nonlinear behavior during the transient growth phase.
A second regime follows, between the rhombus and circle points, where the quadratic relationship shown in panel (d) becomes apparent. The streak, vorticity, and roll intensities decrease but fluctuate around a curve consistent with the quadratic scaling. This behaviour suggests that vorticity continues to feed the rolls, while the rolls attempt to regenerate the streaks, completing the SSP loop. However, the decay dominates and turbulence is not sustained, and the system drifts away from the quadratic relationship. Finally, after the circle point, the flow relaxes into the laminar state. Notably, streaks persist longer than rolls, which agrees with the previous decay-rate analysis and with quenching experiments \cite{Liu_Semin_Klotz_Godoy-Diana_Wesfreid_Mullin_2021}. This delayed decay reflects the lift-up effect once the rolls have vanished. 
In summary, the nonlinear character of the transient growth emerges when the initial perturbation amplitudes exceed the threshold required to trigger nonlinear mechanisms, while lower amplitudes remain within a nearly linear evolution.
A few runs, however, do not exhibit this pattern.

\begin{figure}[h]
    \centering
    \includegraphics[scale=0.5]{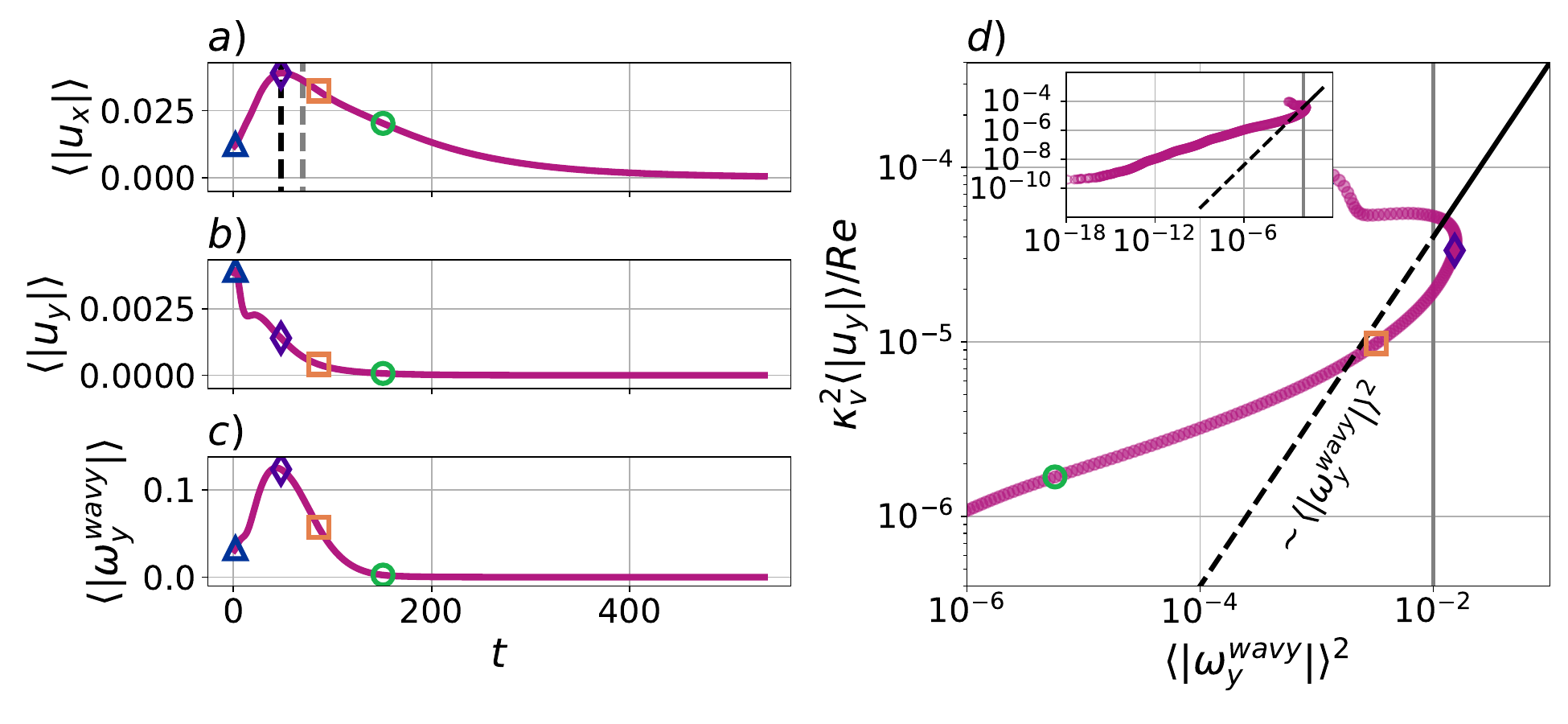}
    \caption{Same as Fig.~\ref{Fig6_uy_wy2_turb}, but for simulation L3 ($Re = 806$), with lower initial intensity. The main panel focuses on the early evolution, while the inset shows the full time series, which decays to the laminar state. The marked times are $t = 0$, $t= 47$, $t=85$, and $t=150$ (triangle, rhombus, square, and circle, respectively).}
    \label{Fig8_uy_wy2_L}
\end{figure}

Figure~\ref{Fig8_uy_wy2_L} shows the same analysis for a weakly perturbed run (L3). In this case, wall-normal vorticity (panel c) remains much smaller than in the other cases, and the quadratic trend is essentially absent. This stage, corresponding to the second regime identified in Fig.~\ref{Fig7_uy_wy2_NL} for case NL2, is precisely what differentiates the simulations labeled “L” from those labeled “NL”. Even though vorticity initially grows slightly, rolls do not develop significantly. This indicates that streak waviness is insufficient to regenerate the rolls. The maximum of $\langle |u_x|\rangle$ occurs close to the linear optimal time (Eq.~(\ref{eqn:topt}) Appendix \ref{Appendix 2}), which points to a fully linear behaviour.

\begin{figure}[H]
    \centering
    \includegraphics[scale=0.5]{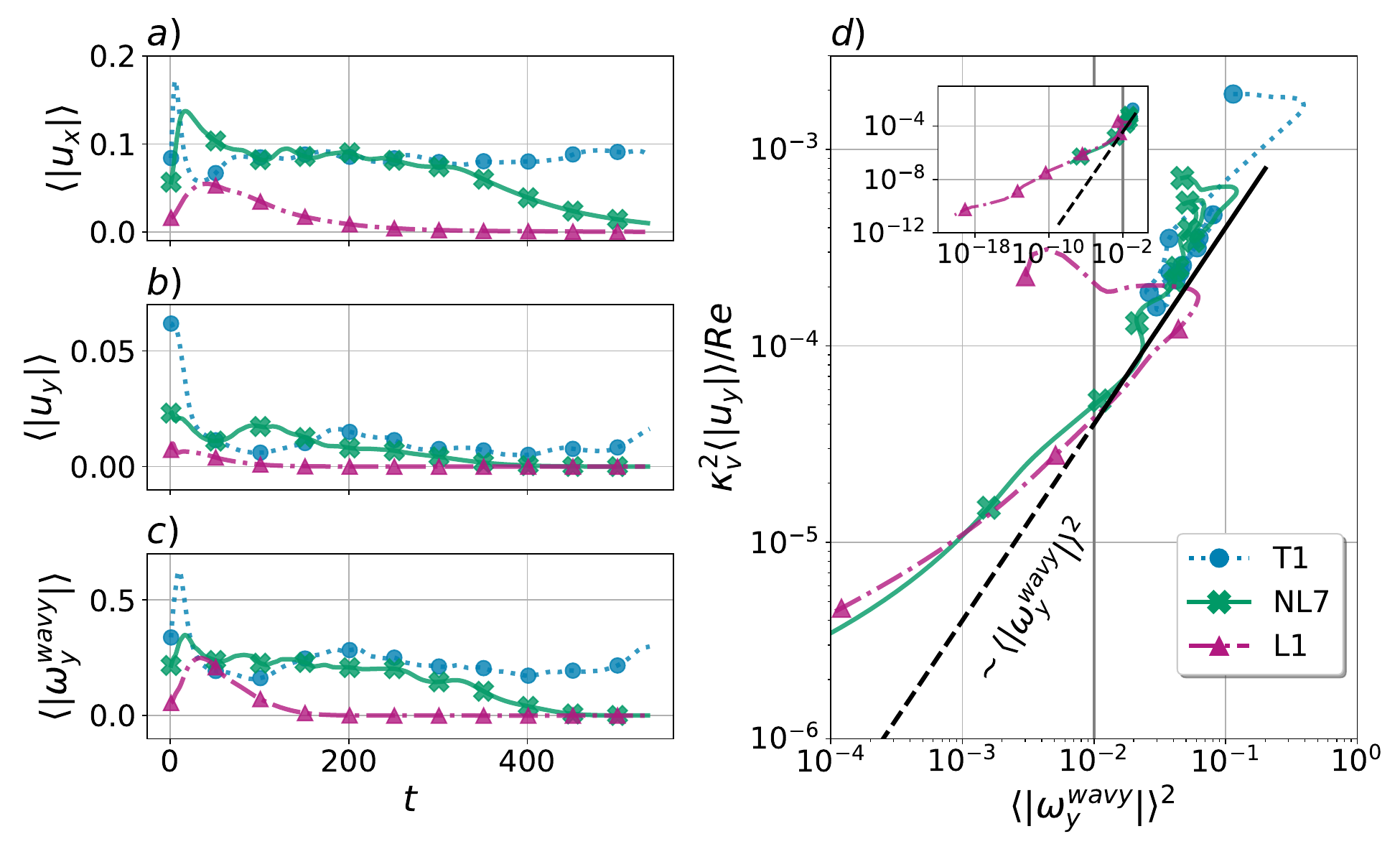}
    \caption{Left panels show the temporal evolution of $\langle |u_x|\rangle$, $\langle |u_y|\rangle$, and $\langle |\omega_y^{wavy}|\rangle$ for simulations T1, NL7, and L1 ($Re = 625$). Panel (d) displays $\kappa_v^2\langle |u_y|\rangle /Re$ vs. $\langle |\omega_y^{wavy}|\rangle^2$ for these runs, focusing on the quadratic region. The inset shows the complete time history, including laminar decay when applicable.}
    \label{Fig9_uy_wy2_Re625}
\end{figure}

\begin{figure}[H]
    \centering
    \includegraphics[scale=0.5]{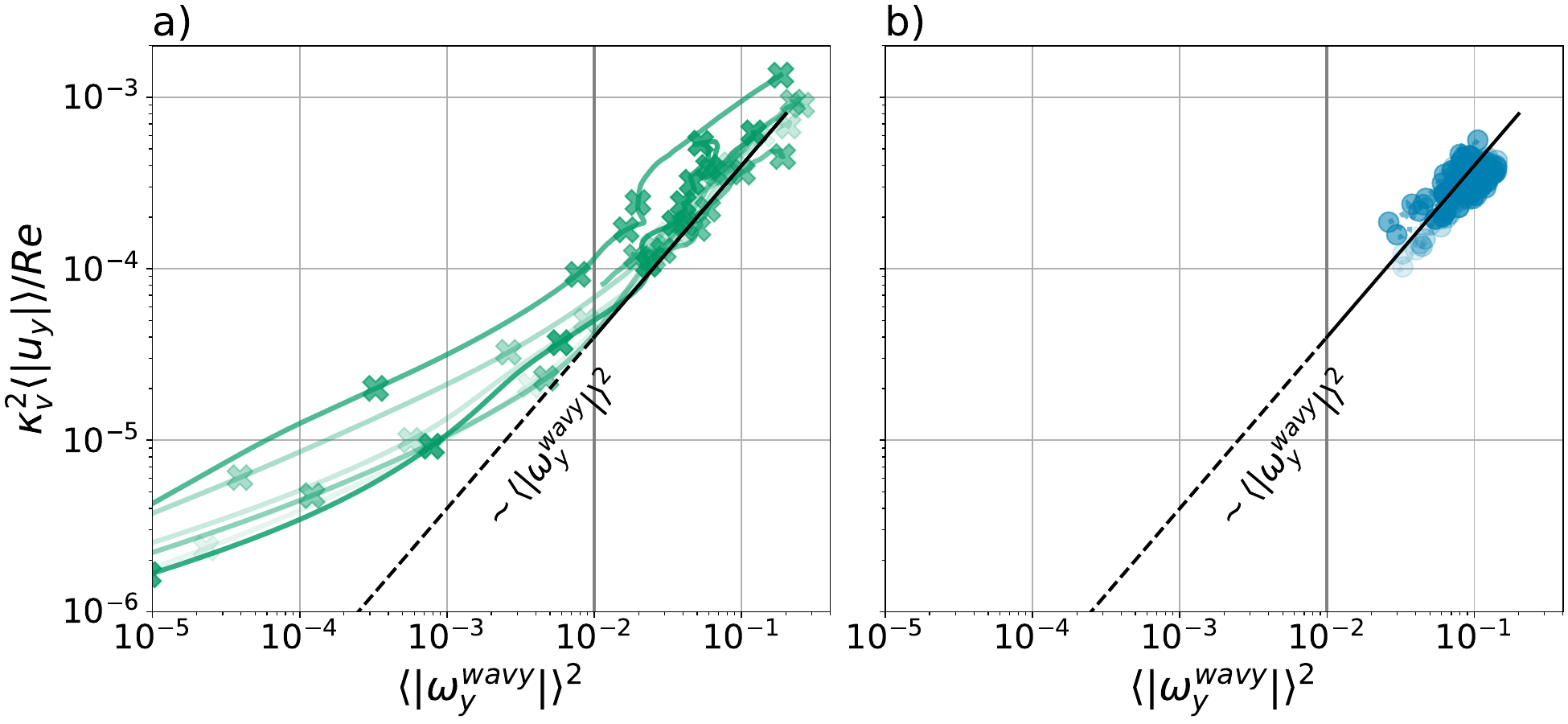}
    \caption{Initial part of $\kappa_v^2\langle |u_y|\rangle /Re$ vs. $\langle |\omega_y^{wavy}|\rangle^2$ for NL1–NL7 and T1–T26 simulations. Green solid lines with crosses correspond to NL runs (from $t=15$) in panel (a). Blue dotted lines with circles correspond to T runs (from $t=100$) in panel (b). Opacity varies between runs within each group from lighter (lower opacity) to darker (higher opacity) as the simulation index in table \ref{table1} increases. The black solid line indicates the range where the quadratic scaling holds; the dashed line shows the same relationship outside this range. All turbulent states lie on this nonlinear quadratic scaling between rolls and waviness.}
    \label{Fig10_uy_wy2_all}
\end{figure}

Figure~\ref{Fig9_uy_wy2_Re625} compares the three behaviours (L, NL, and T) at the same Reynolds number ($Re=625$), for runs T1, NL7, and L1. The left panels display the time series of the streaks (a), rolls (b), and waviness (c). The right panel (d) shows the $\kappa_v^2\langle |u_y|\rangle /Re$ vs. $\langle |\omega_y^{wavy}|\rangle^2$ relationship, zoomed in on the quadratic region. The black solid line highlights the region where the scaling holds, while the inset shows the full temporal evolution, including the eventual decay to zero. The fluctuation levels consistent with the quadratic scaling suggested by Eq. \ref{V_waleffe} in the NL run match those observed in the turbulent run, which is made even clearer in Fig.~\ref{Fig10_uy_wy2_all}. This figure gathers all T and NL simulations, showing that turbulent states are compatible with the fluctuating regime of the NL runs, provided the waviness is sufficiently large. The correlation range is approximately $[4\times 10^{-5},8\times 10^{-4}]$ for $\kappa_v^2\langle |u_y|\rangle /Re$ and $[1\times 10^{-2},2\times 10^{-1}]$ for $\langle |\omega_y^{wavy}|\rangle^2$, although these should not be interpreted as strict thresholds. The black solid line marks this correlation range and the dashed line extends it outside this range. This scaling holds for all Reynolds numbers considered in this study.

\section{Conclusions}\label{Conclusions}

In plane Couette–Poiseuille flow, as in other wall-bounded shear flows, turbulence can be sustained through the interaction between streaks and rolls. This work provides a quantitative and systematic analysis of streak waviness dynamics during the transition to turbulence, demonstrating that streak waviness is a key ingredient in the self-sustaining process (SSP). 

We performed direct numerical simulations over a range of Reynolds numbers and initial conditions. Proxies were defined for the streaks ($\langle |u_x|\rangle$), the rolls ($\langle |u_y|\rangle$), and the streak waviness ($\langle |\omega_y^{wavy}|\rangle$), with spatial averages taken over the entire domain.

In all investigated cases, no significant differences were observed between the types of initial conditions used in this work. The dynamics leading to either laminar decay or sustained turbulence appears to be largely independent of the form of the initial perturbation, but it is strongly sensitive to its intensity, as one would expect. Streamwise rolls were artificially introduced to help optimize transient growth \cite{Cherubini2011,Pringle2012}, but these rolls were constructed in an \textit{ad hoc} manner and could likely be improved in future studies. Regardless of the initial condition, the streaks, rolls, and waviness first grow before either decaying to a laminar state or persisting as turbulence. When the Reynolds number is high and the initial perturbation is large, the system evolves toward a turbulent steady state. In contrast, for weaker perturbations, the flow relaxes back to the laminar state. In this laminar regime, the streaks and rolls decay exponentially, with the rolls decaying approximately twice as fast as the streaks \cite{Liu_Semin_Klotz_Godoy-Diana_Wesfreid_Mullin_2021}.

The central result of this study is that the roll amplitude $\langle |u_y|\rangle$ follows a quadratic relationship with the streak waviness $\langle |\omega_y^{wavy}|\rangle$, provided that the waviness exceeds a threshold of approximately $10^{-2}$. 
More precisely, $\kappa_v^2\langle |u_y|\rangle /Re$ is proportional to $\langle |\omega_y^{wavy}|\rangle^2$, with $\kappa_v \simeq 4.4$ and a proportionality constant $\sigma_v \approx 4 \times 10^{-3}$ in all cases.
This confirms quantitatively that Eq.~(\ref{V_waleffe}), originally derived by Waleffe for plane Couette flow, also holds in Couette–Poiseuille flow. 

This quadratic relationship is observed throughout the turbulent regime (after the brief initial transient) and during the nonlinear decaying phase when the initial perturbation is sufficiently large, even when the final state remains laminar.
The scaling holds for Reynolds numbers between 500 and 940 and is largely independent of the initial perturbation shapes considered.
Conversely, when the initial perturbation is weak, waviness remains below the $10^{-2}$ threshold, the quadratic scaling is absent, and the flow stays laminar.

Earlier experiments by Duriez \textit{et al.} \cite{duriez2009self} provided qualitative evidence for this mechanism, while more recent measurements \cite{Liu2024} offered clearer observations of the lift-up effect and semi-quantitative agreement with Waleffe's model. These experiments successfully captured the link between streak amplitude and waviness through Eq.~(\ref{U_waleffe}) but could not resolve the quadratic relationship between rolls and waviness from Eq.~(\ref{V_waleffe}), which motivated the present numerical study. Unlike 2D PIV measurements, which provide local information, our numerical approach yields global amplitudes, allowing us to probe more clearly the nonlinear regime where the SSP is activated.

The proxy variables defined in this study quantify the amplitudes of rolls, streaks, and waviness, and are intentionally simple and broadly applicable, yet they successfully capture the key nonlinear link between streak waviness and the instability that regenerates the rolls. To our knowledge, this relationship has not been clearly quantified before. While previous experimental works \cite{Liu2024} could identify the lift-up mechanism, they were not able to evidence the nonlinear waviness–roll coupling that is shown here. The present analysis therefore provides a new quantitative perspective on how waviness contributes to the self-sustaining process.

Despite this success, the simplicity of the proposed variables also imposes certain limitations. In particular, when the flow becomes turbulent, the lift-up process tends to collapse into a single point in the phase space representation, suggesting that these variables may not clearly reveal the detailed dynamics of the cycle. Extending the analysis to other terms in Waleffe’s model requires careful treatment of the relationships between the variables to ensure consistency.

The relationship observed in turbulent states suggests that this aspect of the SSP can be effectively captured using only two variables—rolls and waviness—reinforcing the physical insight underlying Waleffe's reduced model.
Furthermore, it is interesting to consider that large-scale flows \cite{benavides2025,ciola2025large,gome2023patterns_part1}, which primarily arise from spatial inhomogeneities inducing normal vorticity, may amplify the effects discussed in this work.
Future work could examine additional nonlinear couplings, explore the role of different spatial modes, or test the robustness of this scaling in other geometries and boundary conditions.

\begin{acknowledgments}

We would like to acknowledge the valuable discussions with Y. Duguet, T. Liu, P. Mininni, L. Tuckerman, D. Barkley, F. Pugliese and B. Español. M.E. acknowledges funding from CNRS. M.E.~and P.D.~acknowledges financial support from the following grants: PIP Grant No.~11220200101752, 
PICT ANPCyT Grant No.~2018-4298, UBACyT Grant No. 20020220300122BA and Redes de Alto Impacto REMATE from Argentina.

\end{acknowledgments}

\newpage
\appendix
\section{Transition to turbulence exploration}\label{Appendix 1}
\begin{figure}[H]
    \centering
    \includegraphics[scale=0.7]{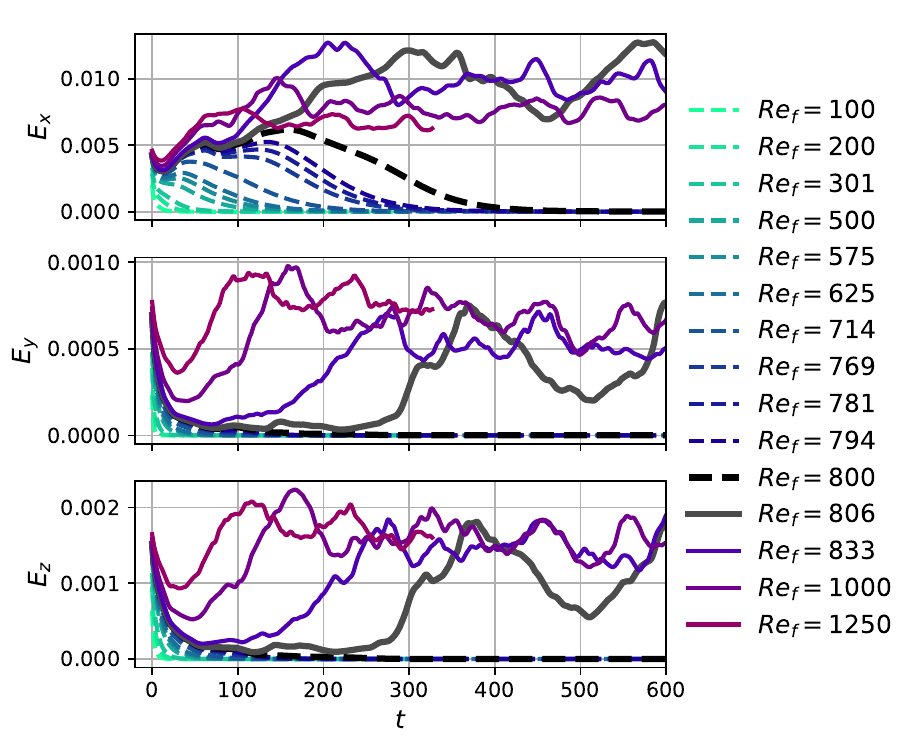}
    \caption{Evolution of velocity component energies over time in the \textit{quench} simulations. The upper panel show the streamwise energy $E_x$, the middle panel show the wall-normal energy $E_y$ and finally the spanwise energy $E_z$ is shown in the lower panel. The runs closest to the critical Reynolds number are indicated in gray solid line (turbulent) and black dashed line (laminar). Additionally, the runs that reach a turbulent regime are represented with solid lines, while the laminar ones are shown with dashed lines.}
    \label{Fig11_energies_quench}
\end{figure}

\noindent With the aim of finding the critical Reynolds number above which the system becomes turbulent and this state is sustained, a parametric exploration was prepared. The simulations performed with this objective, imitate quench experiments. First of all a fully developed turbulent regime at a high $Re=2000$ was reached. Then, the viscosity coefficients were adjusted to define the final Reynolds numbers ($Re_f$) values and were allowed to evolve until reaching the new steady state. The temporal evolution of the fluctuation energy associated with the streamwise velocity component is shown in Fig.~\ref{Fig11_energies_quench}, where this streak energy is defined as
\begin{equation}
    E_x = \frac{1}{l_xl_yl_z}\int_0^{l_x} \int_0^{l_y} \int_0^{l_z} \left(U_x(x,y,z) - \langle U \rangle_{x,z}(y)\right)^2 dxdydz,
    \label{eq:Ex}
\end{equation}
where $l_x$, $l_y$, and $l_z$ are the nondimensional domain lengths defined from $L_x = l_x L_0$, $L_y = l_y L_0$, and $L_z = l_z L_0$. The turbulent and laminar runs that are near the transitional value, are marked in gray solid line ($Re = 806$) and black dashed line ($Re = 800$) respectively. This study suggested that the parameter range explored contains the transitional range. Finally it is to be noted that the energies of the spanwise and wall-normal velocity components follow the same behavior, which is consistent with the fact that both of them are equivalent for measuring the rolls intensities.

\newpage
\section{Linear transient growth}\label{Appendix 2}
\noindent Plane Couette-Poiseuille flow (CPF) is known to be linearly stable under certain conditions, such as when the upper plate velocity $U_{0}$ exceeds $70\%$ of the centerline velocity~\cite{Potter_1966}. However, infinitesimal perturbations may still undergo significant transient growth due to the non-normality of the linearized Navier--Stokes equation~\cite{bergstrom2005nonmodal}.
In this study, we investigate the transient growth associated with the base flow profile $U_x^{lam}(y)=f_0 Re\,y^2+ (1-f_0 Re)\,y$, for Reynolds number in the range $Re\in [500,1000]$. For each combination of streamwise and spanwise wavenumber, denoted as $k_x$ and $k_z$, respectively, the transient energy growth at time $t$ is quantified by the growth function 
\begin{align}
    G(t) \equiv G(t,k_x,k_z) = \max_{\boldsymbol{q}_0 \ne 0} \frac{\| \boldsymbol{q}(y,t) \|^2}{\| \boldsymbol{q}_0\|^2},
    \label{eqn:growthfunction}
\end{align}
which represents the maximum possible energy amplification of kinetic energy of an initial perturbation. Here $\boldsymbol{q}=[u,v,w]$ denotes the velocity perturbation vector.
The maximum growth $G_{\max}$ at $t=t_{\max}$ is defined as $   G_{\max}\equiv G_{\max}(k_x,k_z) =\max_{t\ge 0} G(t,k_x,k_z)$, and the optimum energy growth $G_{\rm opt}$ at $t=t_{\rm opt}$ is defined as $  G_{\text{opt}}=\max_{k_x,k_z} G_{\max} \left( k_x,k_z \right)$. The procedure for computing the growth function follows the approach outlined in \cite{schmid2012stability}.
\begin{figure}[H]
    \centering
    \includegraphics[scale=1.2]{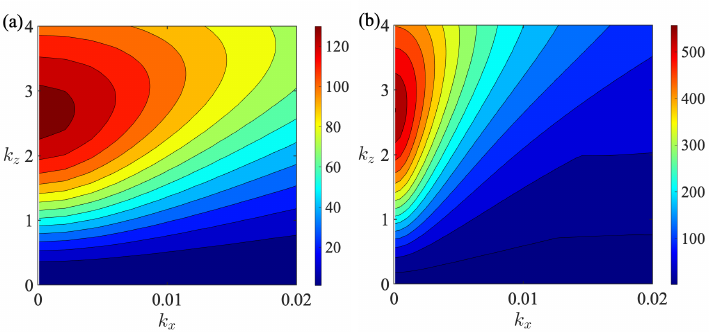}
    \caption{Contours of the maximum gain $G_{\max}$ in the $(k_x,k_z)$-plane for $Re=$ (a) 500 and (b) 1000. The optimal amplifications $G_{\operatorname{opt}}$ are 134.13 and 557.10, occurs at $(\alpha_{\operatorname{opt}},\beta_{\operatorname{opt}})=$ (0,2.653) and (0,2.685), respectively.}
    \label{Fig12_energy_growth}
\end{figure}
Figures~\ref{Fig12_energy_growth} (a) and~\ref{Fig12_energy_growth}(b) display the contours of the maximum gain $G_{\max}$ in the $(k_x,k_z)$-plane for Reynolds numbers $Re=500$ and $1000$, respectively. Clearly, significant transient growth is observed throughout the $(k_x,k_z)$-plane, even though the flow remains asymptotically stable. The optimal growth $G_{\operatorname{opt}}$ occurs at $k_x=0$ (streamwise-independent perturbation), similar to the case of plane Poiseuille flow~\cite{schmid2012stability}, and already observed experimentally in zero mass flux Couette-Poiseuille flow \cite{Klotz_Wesfreid_2017}. 
Additionally, it is found that for the range of Reynolds numbers considered, the optimal spanwise wavenumber $k_{z\operatorname{opt}}$ is approximately $7.48\times10^{-5}Re+2.614$, compared to the fixed value of $2.04$ in plane Poiseuille flow. Furthermore, the relationships between $G_{\operatorname{opt}}$, $t_{\operatorname{opt}}$ and $Re$ are found as:
\begin{align}
    G_{\operatorname{opt}} &\approx 0.00068 Re^2-0.166Re+53.07,\label{eqn:Gopt}\\
    t_{\operatorname{opt}} &\approx 0.0849Re-1.985. \label{eqn:topt}
\end{align}

\bibliographystyle{apsrev4-2}
\bibliography{references}
\end{document}